\documentstyle{article}
\newcommand   {\etal}    {{\it et~al.}}

\newcommand{\beq}{\begin{equation}}
\newcommand{\eeq}{\end{equation}}
\newcommand{\beqa}{\begin{eqnarray}}
\newcommand{\eeqa}{\end{eqnarray}}
\newcommand{\bea}{\begin{eqnarray}}
\newcommand{\eea}{\end{eqnarray}}

\newcommand   {\ehh}     {\epsilon_{\mbox{\tiny HH}}}
\newcommand   {\ehp}     {\epsilon_{\mbox{\tiny HP}}}
\newcommand   {\epp}     {\epsilon_{\mbox{\tiny PP}}}
\newcommand   {\ns}      {S_N}
\newcommand   {\nx}      {D_N}

\input{psfig}
\begin{document}
\begin{titlepage}

\begin{flushright}
Revised version\\
LU TP 97-17\\
\today\\
\end{flushright}

\vspace{0.8in}

\LARGE
\begin{center}
{\bf Local Interactions and Protein Folding: 
A Model Study on the Square and Triangular Lattices\\}
\vspace{.3in}
\large
Anders Irb\"ack\footnote{irback@thep.lu.se} and 
Erik Sandelin\footnote{erik@thep.lu.se}\\ 
\vspace{0.10in}
Complex Systems Group, Department of Theoretical Physics\\ 
University of Lund,  S\"olvegatan 14A,  S-223 62 Lund, Sweden \\
{\tt http://thep.lu.se/tf2/complex/}\\
\vspace{0.3in}	

Submitted to {\it Journal of Chemical Physics}

\end{center}
\vspace{0.3in}
\normalsize
Abstract:

We study a simple heteropolymer model containing sequence-independent local 
interactions on both square and triangular lattices. Sticking to a two-letter
code, we investigate the model for varying strength $\kappa$ of the local
interactions; $\kappa=0$ corresponds to the well-known HP model
[K.F. Lau and K.A. Dill, {\it Macromolecules} {\bf 22}, 3986 (1989)].
By exhaustive enumerations for short chains, we obtain all 
structures which act as a unique and pronounced energy minimum 
for at least one sequence. We find that the number of such designable 
structures depends strongly on $\kappa$. Also, we find that the number
of designable structures can differ widely for the two lattices at a given 
$\kappa$. This is the case, for example, at $\kappa=0$, which implies that 
the HP model exhibits different behavior on the two lattices. Our findings 
clearly show that sequence-independent local properties of the chains can play
an important role in the formation of unique minimum energy structures.    

\end{titlepage}

\newpage
\section{Introduction}
Natural proteins fold into unique compact structures in spite of
the huge number of possible conformations~\cite{Creighton:93}. It is widely
believed that for most single domain proteins, the native structure
is the global free-energy minimum~\cite{Anfinsen:73}, but the
mechanism that determines the structure is still not understood. It 
is also not known whether the property of having a unique native 
structure is common or rare among random polypeptides. It is therefore 
tempting to find out under which conditions and to what extent unique 
native structures appear in simple heteropolymer models.  

In recent years there has been an increasing interest in simple  
statistical-mechanical models for protein folding~\cite{Karplus:95}. 
Most of the models that have been studied are lattice-based with contact 
interactions only. Such models focus on heterogeneity as the primary 
force which drives the formation of unique native structures, and it has
been found that the degree of heterogeneity plays an important role. 
For example, for the cubic lattice, it has been shown that it is possible to  
design twenty-letter sequences that have unique native structures, 
whereas two-letter sequences with this property seem to be 
rare~\cite{Shakhnovich:93,Yue:95}. However, a limitation of these models is 
that local interactions are neglected, and such interactions might be
important not only for the local structure of the chains, as illustrated by  
a recent study of a simple three-dimensional off-lattice 
model~\cite{Irback:97}. This model, with a simple two-letter code, was 
studied both with and without sequence-independent local interactions. 
In the presence of these interactions, it was shown that it is possible 
to find sequences that have compact, well-defined native structures. 
Without the local interactions, no such sequences were found.      
In this paper we investigate the effects of local interactions 
in some detail in two-dimensional lattice models using a two-letter code.

Our starting point is the HP model of Lau and Dill~\cite{Lau:89}, where 
the monomers are either hydrophobic (H) or hydrophilic/polar (P). This model
contains no explicit local interactions, but the underlying lattice can be 
thought of in terms of local interactions. In order to study the influence of 
the lattice, we compare the behavior of the model on the square and triangular 
lattices. Our calculations are based on exhaustive enumerations of the 
full conformational space, for chains containing up to eighteen monomers on 
the square lattice and up to thirteen monomers on the triangular lattice. 
The triangular lattice has the advantage over the more widely used square 
lattice that it does not exhibit the well-known even-odd problem --- on the 
square lattice it is impossible for two monomers at an even distance along 
the sequence to form a nearest-neighbor contact.  

The properties of the HP model on the square lattice are known in some detail 
from previous studies~\cite{Dill:95}. In particular, it has been shown that 
about 2\% of all HP sequences possess non-degenerate ground states on this 
lattice~\cite{Chan:91,Chan:94}. It turns out that such sequences are much more
rare on the triangular lattice. There is, for example, no 13-mer  
with non-degenerate ground state on this lattice. 

Having seen this, we turn to a simple extension of the HP model which contains 
explicit sequence-independent local interactions. We study this model for 
varying strength $\kappa$ of the local interactions, focusing on the set of 
all ground states that are non-degenerate and separated from the rest of 
the states by a sufficiently large energy gap. Sequences having such ground 
states can design the corresponding structures, and the number of sequences 
that can design a given structure will be called the designability of this 
structure~\cite{Li:96}. Every structure that can be designed by at least one 
sequence will be called designable.  

We find that the number of designable structures is strongly 
$\kappa$-dependent, and that it can differ widely for the two lattices at a 
given $\kappa$. The difference is particularly large at $\kappa=0$, which
corresponds to the HP model. However, the maximum numbers of designable 
structures for the two lattices are comparable, and for both lattices there 
is a pronounced peak in the number of designable structures at $\kappa=-0.5$.  
At this $\kappa$ the typical designable structure is compact with many 
turns. Focusing on maximally compact structures, 
we study the designability of individual structures at this $\kappa$.
We find that the designability tends to be much 
higher on the square lattice. On this 
lattice we find that there are certain compact structures which can be 
designed by a very large number of sequences. This finding is in line with 
the results of Li \etal~\cite{Li:96}. However, it is important to note that 
the results for the triangular lattice are different, which indicates that 
the emergence of such structures to some extent is related to the 
even-odd problem.   

Our results clearly show that sequence-independent local interactions can play 
an important role in the formation of unique minimum energy structures. 
Although our study is confined to two dimensions, we expect this to hold in 
three dimensions as well. In fact, one may expect such interactions to be 
even more important in three dimensions, where the flexibility of the chains 
is greater.    
\section{Methods}
The chains studied are linear and self-avoiding, and contain two monomer 
types, H and P. A sequence is specified by a choice of monomer types at each 
position on the chain, $\{\sigma_i\}$, where $\sigma_i$ takes the values H 
and P and $i$ is a monomer index. A structure is specified by a set of 
coordinates for all the monomers, $\{{\bf x}_i\}$, and the bend angle formed
by sites ${\bf x}_{i-1}$, ${\bf x}_i$ and ${\bf x}_{i+1}$ will be denoted
by $\theta_i$. The energy of a structure is given by 
sequence-independent local interactions and sequence-dependent
nearest-neighbor contact interactions,
\bea
E&=&\kappa E_L + E_G\label{energy}\\
E_L&=&2\sum_{i=2}^{N-1}(1-\cos\theta_i)\label{el}\\
E_G&=&
\sum_{1\le i<j\le N}\epsilon_{\sigma_i\sigma_j}\Delta({\bf x}_i-{\bf x}_j)
\label{eg}\eea
where $\Delta({\bf x}_i-{\bf x}_j)=1$ if ${\bf x}_i$ and ${\bf x}_j$ are 
nearest neighbours on the lattice but $i$ and $j$ are not 
adjacent positions along the sequence, and $\Delta({\bf x}_i-{\bf x}_j)=0$
otherwise. The energy depends on three parameters $\ehh$, $\ehp$ and $\epp$ 
which will be held fixed throughout the paper. 
Following Lau and Dill~\cite{Lau:89}, we take 
\beq
\ehh=-1\qquad \ehp=\epp=0
\label{parameters}\eeq 
The remaining parameter $\kappa$ determines the strength of the local 
interactions. For $\kappa=0$ the 
model is identical to the HP model, the energy being given by minus the 
number of topological HH contacts (two monomers $i$ and $j$ are in 
topological contact if $\Delta({\bf x}_i-{\bf x}_j)=1$). 

In our calculations we focus on the energy spectrum. For a given number of 
monomers, $N$, we compute this for all the $2^N$ possible sequences, 
by numerical enumeration of the full conformational space. 
In this way we determine all sequences having a ground state which
is non-degenerate and separated from the rest of the spectrum by 
a sufficiently large energy gap. We say that such a sequence can design 
its ground state structure, and each structure that can be designed by at 
least one sequence will be called designable. The number of designable 
structures will be denoted by $\nx$ for chains with $N$ monomers.

These definitions involve a parameter $\Delta E$; the gap between the ground 
state and the next lowest level is required to be greater than or equal to 
$\Delta E$. The choice of this parameter is somewhat arbitrary. We tested 
several different values and decided to use $\Delta E=1$, corresponding to 
the energy of one HH contact. Small changes of $\Delta E$ leads to 
qualitatively similar results. 

The gap criterion is important when studying general $\kappa$. To 
illustrate this, let us consider $N=13$ chains on the triangular lattice. 
Here one finds that all ground states are degenerate at $\kappa=0$, 
while 4328 of the sequences have non-degenerate ground states at 
small, positive $\kappa$. However, each ground state becomes effectively
degenerate for sufficiently small $\kappa$, since all the gaps vanish in 
this limit. 

Our choice of gap criterion implies that we look for ground states 
corresponding to a single structure on the lattice. For longer chains
it would probably be more relevant to consider the gap between the ground 
state and the lowest of all states with little structural similarity to the
ground state~\cite{Sali:94,Bryngelson:95,Gutin:95b}. In principle, it would 
be more appropriate to formulate the gap criterion in terms of 
a normalized gap $\Delta E/E_\kappa$, $E_\kappa$ being a $\kappa$-dependent 
energy scale. However, the variation of $E_\kappa$ can, for our purposes, be 
neglected in the $\kappa$ region that is of primary interest 
(small and moderate $|\kappa|$).
  
The total number of sequences that can design structures will be denoted by 
$\ns$ for chains with $N$ monomers. In general $\ns$ is greater than $\nx$,
since different sequences may have the same ground state structure. The 
difference between $\ns$ and $\nx$ is particularly large in the trivial limit
$\kappa\to\infty$. In this limit, there is one structure, the rod-like 
structure which minimizes $E_L$, which can be designed by all sequences, 
so $\ns$ is equal to the size of the full sequence space while $\nx=1$. 
In the limit $\kappa\to-\infty$ the 
situation is similar but slightly different. On the triangular lattice 
there is again one structure, a zig-zag pattern, 
which can be designed by all sequences. This structure is the 
unique maximum of $E_L$. On the square lattice there are, by contrast, 
many different structures that maximize $E_L$.   
  
\section{Results}

\subsection{The number of designable structures}

\subsubsection{$\kappa=0$}

We first consider the model in absence of the local interactions ($\kappa=0$). 
Energy gaps can then take integer values only, so, with our choice of 
$\Delta E$, the gap criterion is met by all sequences with non-degenerate 
ground states. Hence, $\ns$ is here simply 
the number of $N$-mers with non-degenerate ground 
states. In Table~\ref{tab:1} we show the quantities $\ns$ and $\nx$ for 
different $N$ on the square and triangular lattices.
\begin{table}[tb]
\begin{center}
\begin{tabular}{|c|c|c|c|c|c|c|} 
\cline{2-7}
\multicolumn{1}{c|}{}   
&\multicolumn{3}{c|}{\it square}
&\multicolumn{3}{c|}{\it triangular}\\
\hline
    & No. of        &       &       & No. of        &       &       \\
$N$ & conformations & $\ns$ & $\nx$ & conformations & $\ns$ & $\nx$ \\
\hline
3   & 2             & 0     & 0     & 3       & 2 & 1 \\
4   & 5             & 4     & 1     & 12      & 2 & 2 \\
5   & 13            & 0     & 0     & 52      & 1 & 1 \\
6   & 36            & 7     & 3     & 228     & 0 & 0 \\
7   & 98            & 10    & 2     & 996     & 0 & 0 \\
8   & 272           & 7     & 5     & 4324    & 0 & 0 \\
9   & 740           & 6     & 4     & 18678   & 0 & 0 \\
10  & 2034          & 6     & 4     & 80345   & 2 & 2 \\
11  & 5513          & 62    & 14    & 344431  & 6 & 6 \\
12  & 15037         & 87    & 25    & 1472412 & 2 & 2 \\
13  & 40617         & 173   & 52   & 6279601 & 0 & 0 \\
14  & 110188        & 386   & 130   &         &   &   \\
15  & 296806        & 857   & 218   &         &   &   \\
16  & 802075        & 1539  & 456   &         &   &   \\
17  & 2155667       & 3404  & 787   &         &   &   \\
18  & 5808335       & 6349  & 1475  &         &   &   \\
\hline
\end{tabular}
\caption{$\protect\ns$ and $\protect\nx$ for different $N$ for the HP model 
($\kappa=0$) on the square and triangular lattices. Also shown are the total 
numbers of conformations for different $N$.}
\label{tab:1}
\end{center}
\end{table}
Our results for $\ns$ on the square lattice can be compared with those 
of Chan and Dill~\cite{Chan:91,Chan:94} and are consistent with these. 
Also shown in Table~\ref{tab:1} are the total numbers of different
conformations, unrelated by simple symmetries, for different $N$. 

From Table~\ref{tab:1} it can be seen that $\ns$ increases roughly linearly 
with the total number of sequences on the square lattice; the fraction of 
sequences having non-degenerate ground states varies between 2.1 and 2.6\% 
for $12\le N\le 18$. Over the same $N$ range the number of designable 
structures satisfies $0.23\ns<\nx<0.34\ns$, so on average each designable 
structure can be designed by 2.9--4.3 sequences. These 
results contrast sharply with those for the triangular lattice, where   
$\ns$ and $\nx$ are much smaller. Also, there is no structure on the 
triangular lattice that can be designed by more than one sequence for 
$4\le N\le 13$. 

\subsubsection{$\kappa\ne0$}

\begin{figure}[tb]
\begin{center}
\vspace{-43mm}
\mbox{\hspace{-31mm}\psfig{figure=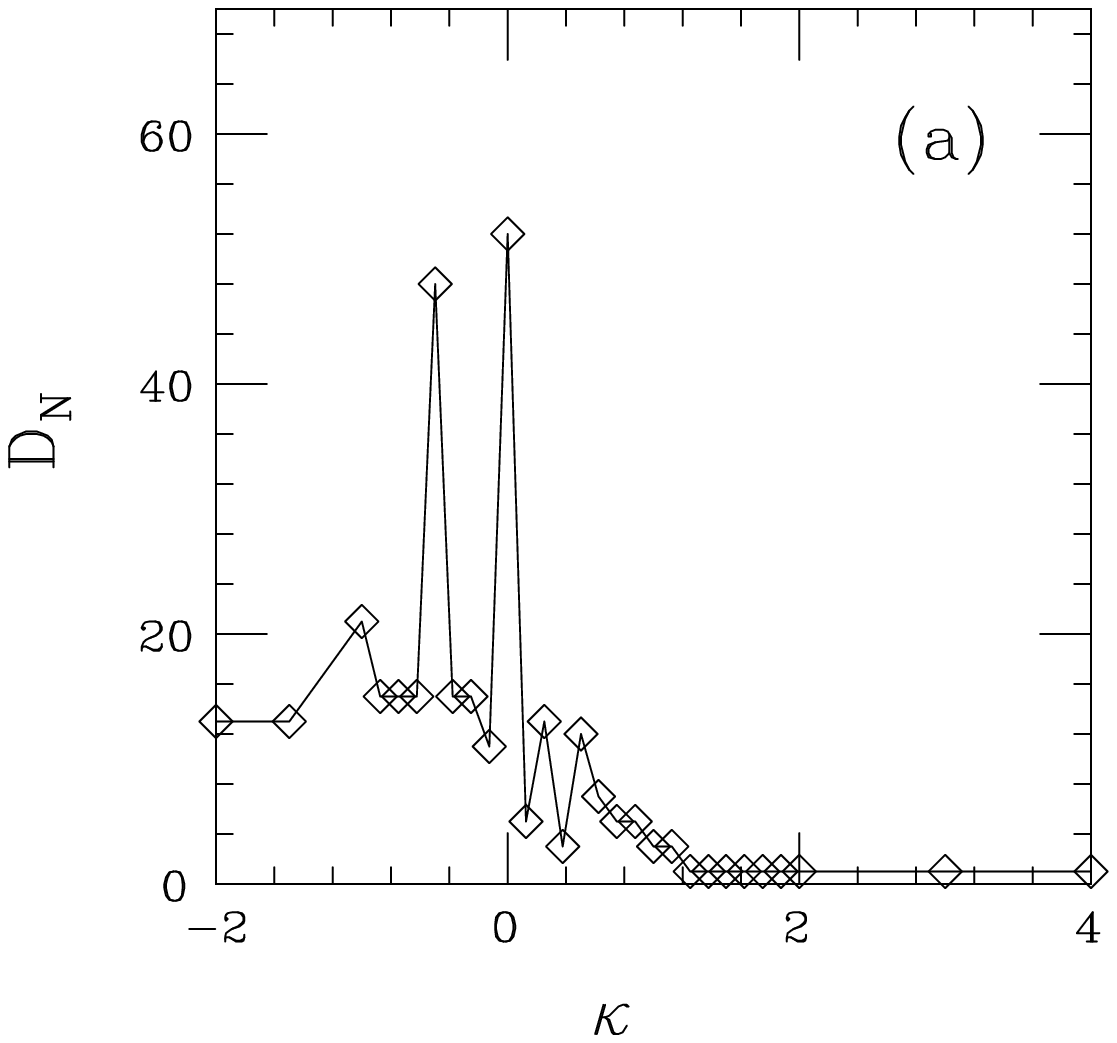,width=10.5cm,height=14cm}
\hspace{-30mm}\psfig{figure=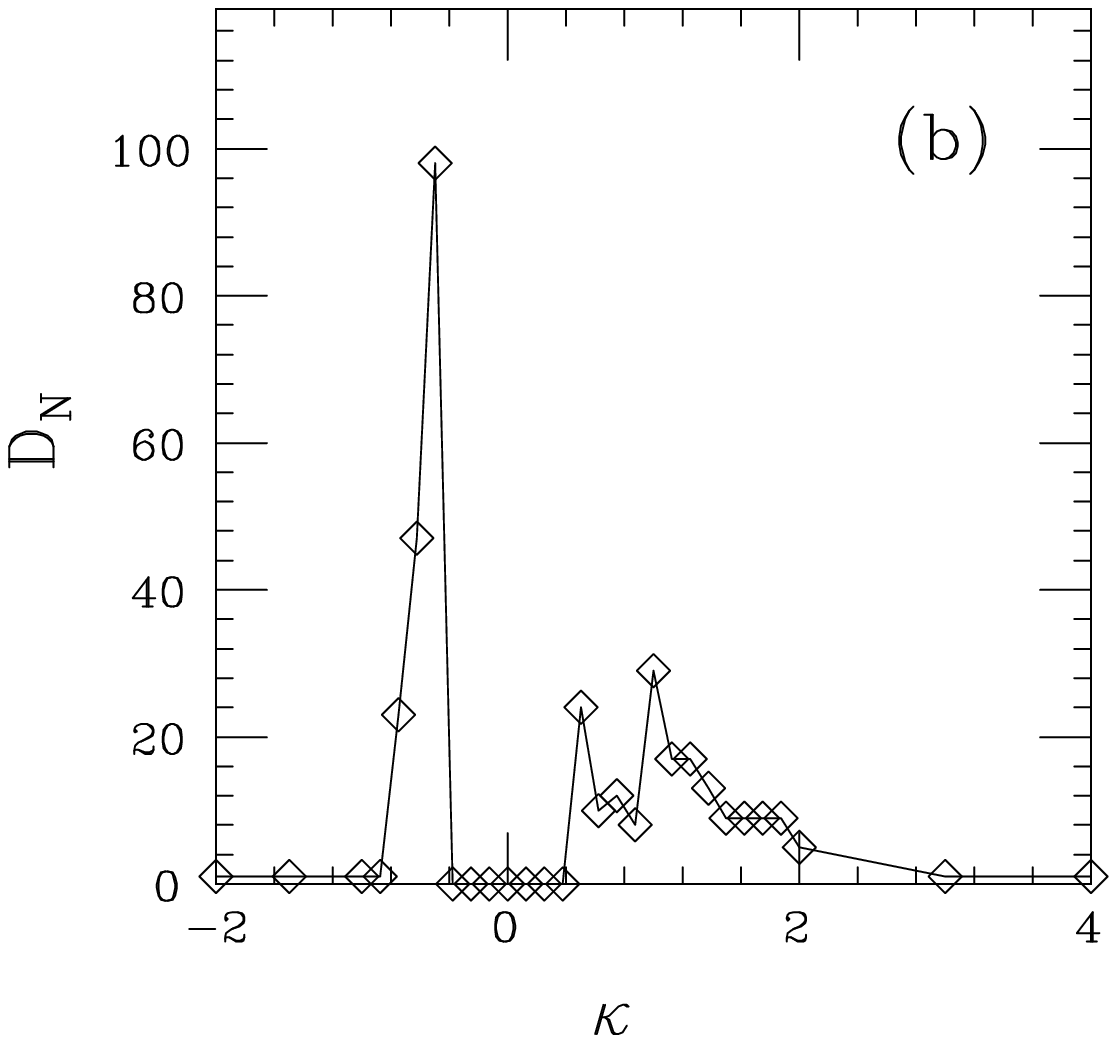,width=10.5cm,height=14cm}}
\vspace{-43mm}
\end{center}
\caption{The number of designable structures, $\protect\nx$, versus $\kappa$ 
for $N=13$, (a) square and (b) triangular lattice.}
\label{fig:1}
\end{figure}

We now turn to general $\kappa$. In Fig.~\ref{fig:1} we show the 
$\kappa$ dependence of $\nx$ for $N=13$. The large-$|\kappa|$ behavior 
of $\nx$ is trivial, as discussed above. However, as can be seen from 
Fig.~\ref{fig:1}, there is a small-$|\kappa|$ region where 
$\nx$ shows an interesting and strong $\kappa$ dependence. 
For both lattices there is a sharp peak at $\kappa=-0.5$. 
For the square lattice there is another, slightly higher peak at $\kappa=0$, 
corresponding to the HP model. Such a peak is missing for the triangular 
lattice, which leads to the big difference in the results for the HP model.
 
\begin{figure}[t]
\vspace{-43mm}
\mbox{
\hspace{-30mm}
\psfig{figure=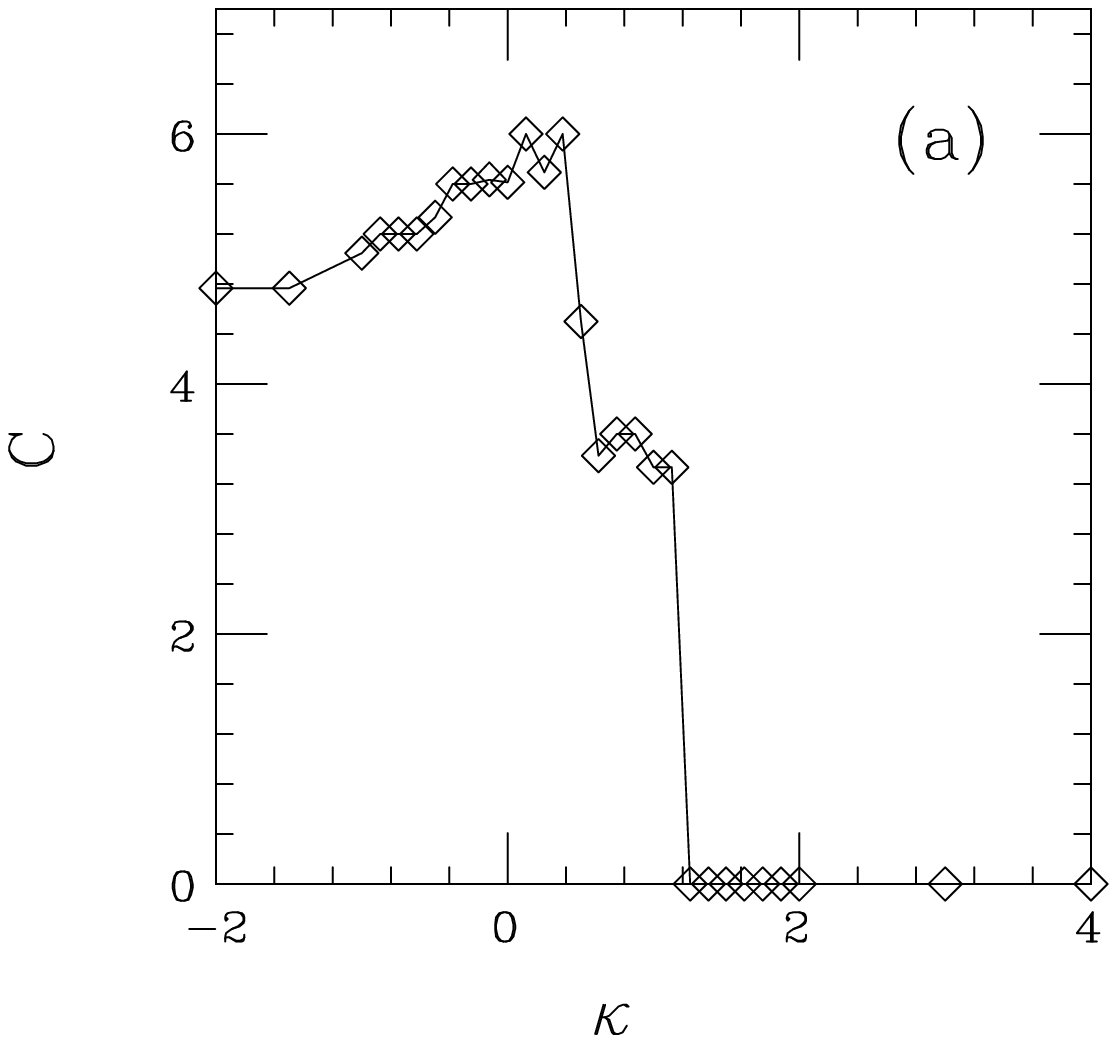,width=10.5cm,height=14cm}
\hspace{-30mm}
\psfig{figure=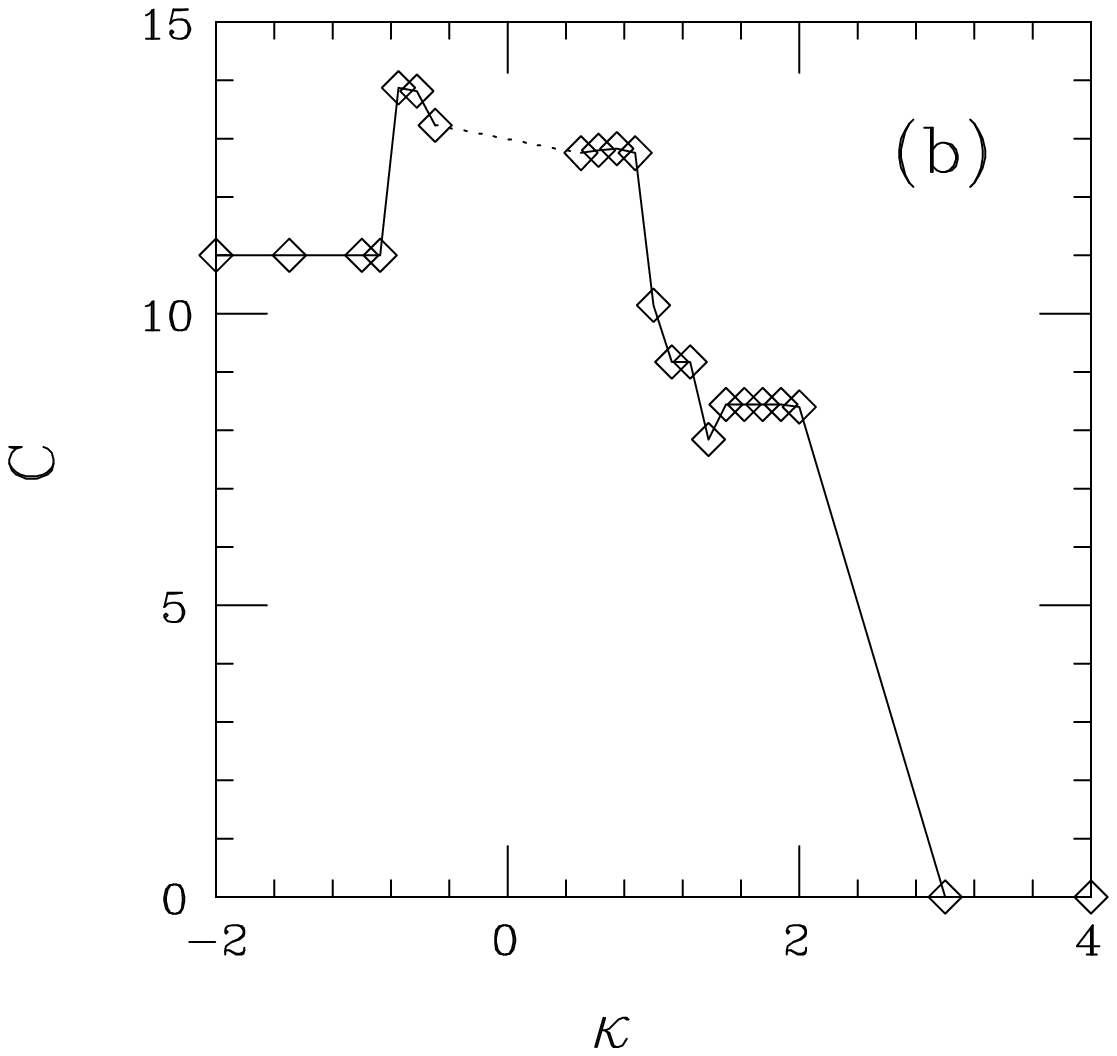,width=10.5cm,height=14cm}
}

\vspace{-75mm}

\mbox{
\hspace{-30mm}
\psfig{figure=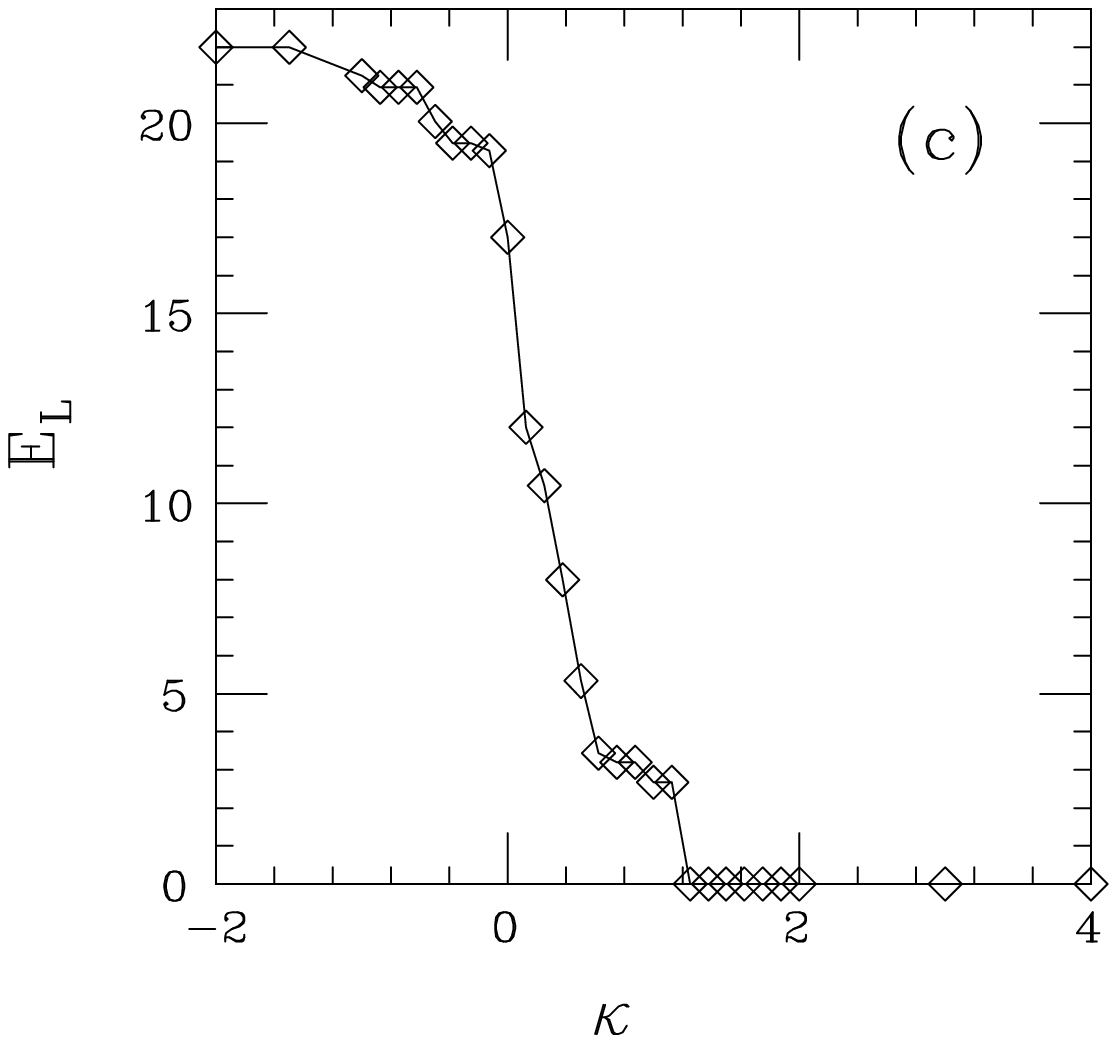,width=10.5cm,height=14cm}
\hspace{-30mm}
\psfig{figure=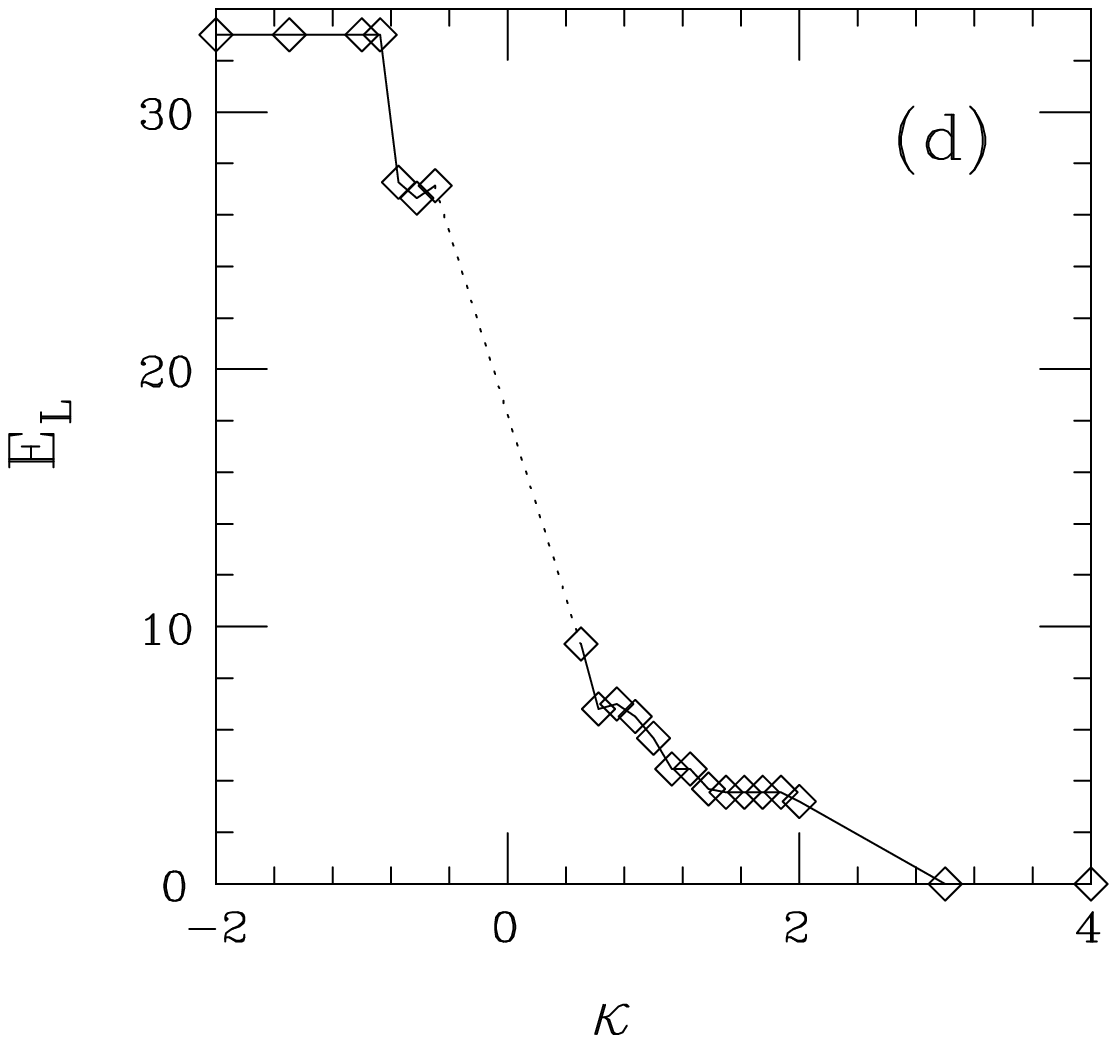,width=10.5cm,height=14cm}
}
\vspace{-43mm}
\caption{Averages of $C$ (the number of topological contacts) and $E_L$ 
(see Eq.~\protect\ref{el}) over the designable structures at different 
$\kappa$ for $N=13$: (a) $C$, square lattice; (b) $C$, triangular lattice;
(c) $E_L$, square lattice; (d) $E_L$, triangular lattice. For the triangular
lattice, there are $\kappa$ values at which no designable structures 
were found (see Fig.~\protect\ref{fig:1}). The lines connecting the data 
points are dotted in this region.}
\label{fig:2}
\end{figure}

In order to study the character of the designable structures, we computed
average values of the total number of topological contacts, $C$, which is a 
measure of compactness, and $E_L$ (see Eq.~\ref{el}). In Fig.~\ref{fig:2}  
these quantities are plotted against $\kappa$, using $N=13$. 
Each data point in this figure is an average over all the 
designable structures at a given $\kappa$. 
For $N=13$, the maximum value of $C$ is 
6 on the square lattice and 14 on the triangular lattice, 
whereas the maximum value of $E_L$ is 22 
on the square lattice and 33 on the triangular lattice.  
From Fig.~\ref{fig:2} it can be seen that $C$ and $E_L$ vary 
rapidly with $\kappa$. 
Both quantities are large at the peaks at $\kappa=-0.5$, 
showing that the typical designable structure at this $\kappa$ 
is compact with many turns. 
   
A closer look at the designable structures at $\kappa=-0.5$ shows
that roughly half of them are maximally compact, in the sense that
they have maximum $C$; this holds for 19 out of 
48 structures on the square lattice, and for 50 out of 98 structures
on the triangular lattice. An example of a designable structure at 
$\kappa=-0.5$ which is not maximally compact is the zig-zag structure 
that maximizes $E_L$ on the triangular lattice. A structure such as this
is less interesting than the maximally compact ones from the 
viewpoint of design. In our study below of the designability of individual 
structures, we focus on maximally compact structures.  

\begin{figure}[tb]
\begin{center}
\vspace{-43mm}
\mbox{\hspace{-31mm}\psfig{figure=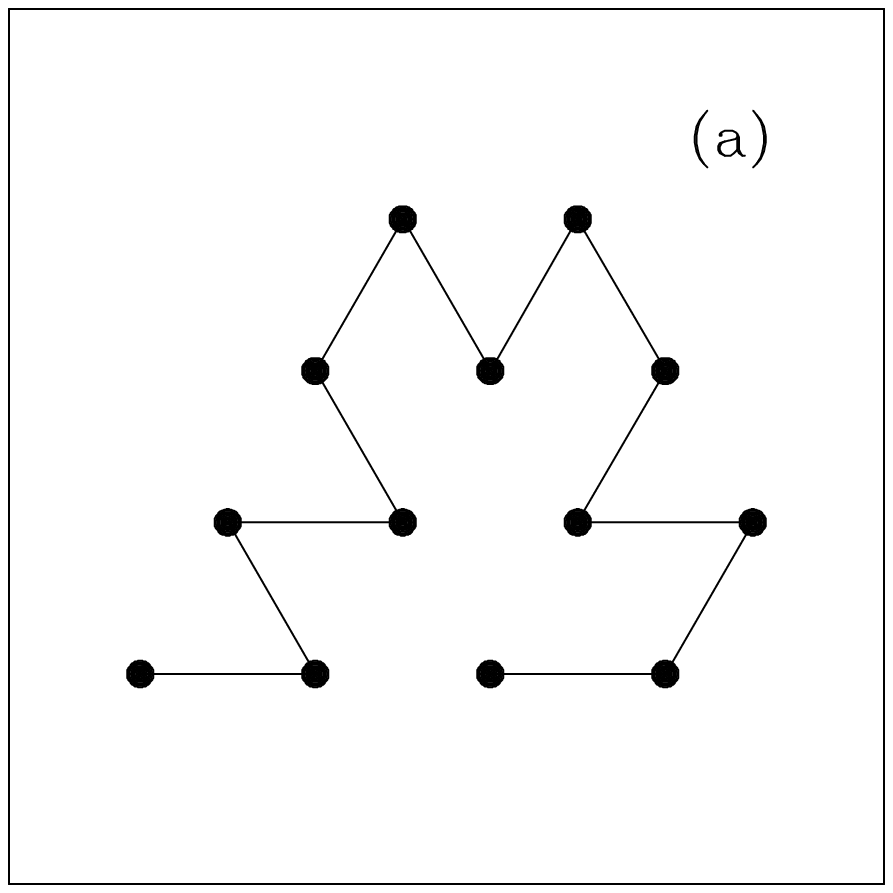,width=10.5cm,height=14cm}
\hspace{-30mm}\psfig{figure=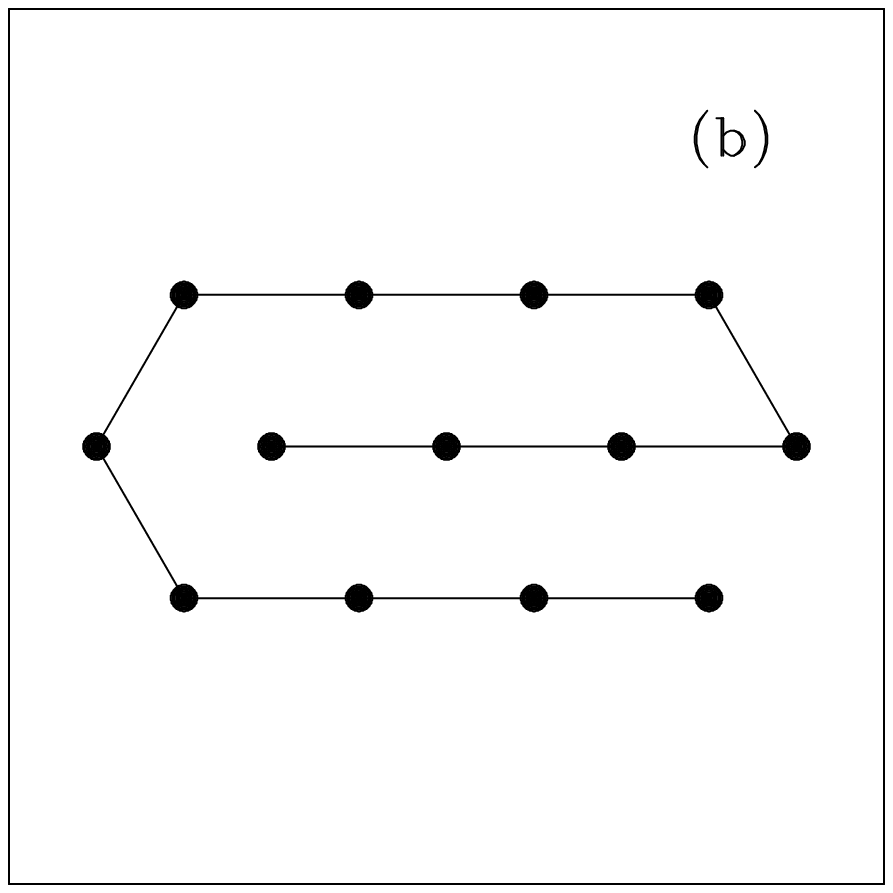,width=10.5cm,height=14cm}}
\vspace{-43mm}
\end{center}
\caption{Examples of designable structures at (a) $\kappa=-0.5$ and 
(b) $\kappa=0.5$.}  
\label{fig:3}
\end{figure}

Local properties of the designable structures, such as $E_L$, are strongly 
$\kappa$-dependent. To illustrate this, we show in Fig.~\ref{fig:3} two 
designable structures on the triangular lattice corresponding to $\kappa=-0.5$ 
and $\kappa=0.5$, respectively. Both these conformations are 
maximally compact. Another important property which they share 
is that they display strong regularities in the local structure,
reminiscent of the secondary structure in real proteins. 
The two conformations differ markedly, however, in the precise form of the 
local structure.   

\subsection{The designability of individual structures}

So far we have classified the structures in a binary way as either 
designable or not. It is also interesting to see to what extent those 
structures that are designable differ in designability, the designability 
being defined as the number of sequences that can design a given structure. 

Recently, Li~\etal~\cite{Li:96} studied the designability of individual 
structures in a HP-like model on the square and cubic lattices. These 
calculations were performed using the restricted conformational space 
consisting of all maximally compact structures, and the energy function 
was given by Eqs.~1--3 with $\ehh=-2.3$, $\ehp=-1$ and $\epp=\kappa=0$. 
Large variations in designability were observed. In particular, it was 
shown that certain structures can be designed by a huge number of sequences.

In order to test the generality of these findings, we have performed 
analogous calculations for our model using $\kappa=-0.5$ and $N=13$. 
The designability was computed for each of the maximally compact  
structures, as discussed above. Let us stress, however, that, in contrast
to Li~\etal~\cite{Li:96}, we base our definition of designability   
upon the full set of all possible structures rather than the subset 
of maximally compact structures. 

Comparing our results for the square and triangular lattices, 
we find that the designability tends to be much higher on the 
square lattice. The average designability is 82.1 for this lattice, 
compared to 4.5 for the triangular lattice. The highest 
designabilities we measured are 590 and 12, respectively, 
for the square and triangular lattices. The variations in designability are,
as in the model studied by Li~\etal~\cite{Li:96}, very large
on the square lattice. 
 
The observed differences in designability for the two lattices are striking,
but may be not surprising in view of the even-odd problem. The fact that
certain contacts cannot be formed on the square lattice tends, for a given
structure, to increase the degree of degeneracy with respect to the sequence 
degrees of freedom.
 
\section{Summary and discussion}

We have studied ground state properties of a simple heteropolymer model 
containing sequence-independent local interactions. By enumeration of 
the full conformational space for short chains, we calculated the degeneracy
of the ground state and the gap to the next lowest level for all possible 
sequences. In this way we obtained all structures that are designable.  
Our results show that the number of designable structures 
depends strongly on the strength $\kappa$ of the local interactions. 
Furthermore, we have seen that the behavior on the square and triangular 
lattices can be very different at a given $\kappa$. An example of 
this is the HP model ($\kappa=0$). As our study of non-zero $\kappa$ shows, 
the difference in behavior of the HP model on the two lattices can 
to a certain degree be compensated for by introducing the local 
interactions. However, there seems to be at least one important 
difference between the properties on the two lattices. Our study of the
designability of maximally compact structures shows that this
tends to be much higher on the square lattice. Furthermore, on this 
lattice we find that, as in the model studied by Li~\etal~\cite{Li:96}, there 
are certain compact structures that can be designed by a huge number of 
sequences. No such structures were found on the triangular lattice.
     
The effects of local interactions were recently studied in a simple 
three-dimensional off-lattice model with two-letter code~\cite{Irback:97}.
Our findings are in good agreement with the results of this study, but may at
sight seem to contradict the results obtained by Govindarajan and 
Goldstein~\cite{Govindarajan:95}. It should therefore be stressed that 
the system studied by Govindarajan and Goldstein is very different from ours. 
They studied the behavior of optimized sequences in a model with a 
much higher degree of heterogeneity, in which the local interactions are 
sequence-dependent. 

Starting with the work of Flory~\cite{Flory:56}, 
there have been a number of studies of homopolymer models containing 
local interactions similar to those we have considered here. 
From the viewpoint of protein folding, the model studied by 
Kolinski~\etal~\cite{Kolinski:86a,Kolinski:86b} appears particularly 
interesting. This model contains, in addition to the local interactions, 
also attractive nearest-neighbor contact interactions.   
Its phase diagram for $\kappa>0$ exhibits coil and globule phases, 
as well as a folded low-temperature phase~\cite{Doniach:96,Bastolla:97}. 
The behavior of this model for $\kappa<0$ has to our knowledge not been
investigated.

The HP model, without local interactions, has recently been utilized by 
Camacho and Schanke~\cite{Camacho:97} in order to study the role of crosslinks 
in polymers. There have been several recent studies of this important issue 
based on Gaussian models~\cite{Solf:95,Bryngelson:96,Kantor:96}. 
Camacho and Schanke took a different approach and examined 
the zero temperature limit of the HP model, where the surviving structures 
are those that have the maximum number of HH links. Their study was carried 
out using the square lattice, and it would be interesting to see to 
what extent the results remain unchanged on the triangular lattice.      
The behavior of the HP model on the triangular lattice has been studied
previously in other contexts,
recently by Seno \etal~\cite{Seno:96}.

\section*{Acknowledgements}

We would like to thank Carsten Peterson, Frank Potthast and Ola Sommelius 
for helpful comments on the manuscript, and Michele Vendruscolo for making 
unpublished data available to us.   

\newpage

\newcommand  {\Biopol}   {{\it Biopolymers\ }}
\newcommand  {\BC}       {{\it Biophys.\ Chem.\ }}
\newcommand  {\BJ}       {{\it Biophys.\ J.\ }}
\newcommand  {\COSB}     {{\it Curr.\ Opin.\ Struct.\ Biol.\ }}
\newcommand  {\EL}       {{\it Europhys.\ Lett.\ }}
\newcommand  {\JCC}      {{\it J.\ Comp.\ Chem.\ }}
\newcommand  {\JCoP}     {{\it J.\ Comp.\ Phys.\ }}
\newcommand  {\JCP}      {{\it J.\ Chem.\ Phys.\ }}
\newcommand  {\JMB}      {{\it J.\ Mol.\ Biol.\ }}
\newcommand  {\JP}       {{\it J.\ Phys.\ }}
\newcommand  {\JPC}      {{\it J.\ Phys.\ Chem.\ }}
\newcommand  {\JSP}      {{\it J.\ Stat.\ Phys.\ }}
\newcommand  {\Mac}      {{\it Macromolecules\ }}
\newcommand  {\MC}       {{\it Makromol.\ Chem.,\ Theory Simul.\ }}
\newcommand  {\MP}       {{\it Molec.\ Phys.\ }}
\newcommand  {\Nat}      {{\it Nature}}
\newcommand  {\NP}       {{\it Nucl.\ Phys.}}
\newcommand  {\Pro}      {{\it Proteins:\ Struct.\ Funct.\ Genet.\ }}
\newcommand  {\ProSci}   {{\it Protein\ Sci.\ }}
\newcommand  {\Pa}       {{\it Physica\ }}
\newcommand  {\PL}       {{\it Phys.\ Lett.\ }}
\newcommand  {\PNAS}     {{\it Proc.\ Natl.\ Acad.\ Sci.\ USA\ }}
\newcommand  {\PR}       {{\it Phys.\ Rev.\ }}
\newcommand  {\PRL}      {{\it Phys.\ Rev.\ Lett.\ }}
\newcommand  {\PRS}      {{\it Proc.\ Roy.\ Soc.\ }}
\newcommand  {\Sci}      {{\it Science\ }}
\newcommand  {\ZP}       {{\it Z.\ Physik\ }}


\end{document}